# Complex magnetic structure in Ba$_5$Ru$_3$O$_{12}$ with isolated Ru$_3$O$_{12}$-trimer

T. Basu[1*], F. Y. Wei[2], Q. Zhang[3], Y. Fang[2], and X. Ke[1]

[1]Department of Physics and Astronomy, Michigan State University, East Lansing, Michigan 48824, USA

[2]Jiangsu Laboratory of Advanced Functional Materials, Department of Physics, Changshu Institute of Technology, Changshu 215500, China

[3]Neutron Scattering Division, Oak Ridge National Laboratory, Oak Ridge, Tennessee 37831, USA

*tathamaybasu@gmail.com

**Abstract**

We report detailed magnetic, transport, heat-capacity, and neutron diffraction measurements of Ba$_5$Ru$_3$O$_{12}$, a compound consisting of isolated Ru$_3$O$_{12}$ trimers. We show that this system develops long-range antiferromagnetic ordering at $T_N$ ~ 60 K without structural distortion or metal-insulator-type transition, which is in sharp contrast to other Barium Ruthenate trimer systems such as 9R-BaRuO$_3$ and Ba$_4$Ru$_3$O$_{10}$. A complex magnetic structure is revealed which is attributable to the magnetic frustration due to competing exchange interactions between Ru ions on different crystallographic sites within the Ru$_3$O$_{12}$ trimer and different degree of orbital hybridization on different Ru-sites. We have also investigated the magnetic structure of Ba$_4$Ru$_3$O$_{10}$ for comparisons.

## I. Introduction

The interplay between electronic correlation of extended 4$d$-orbital, crystal-field effect and strong spin-orbit coupling in Ruthenates yields a rich variety of physical properties, such as superconductivity, Mott-insulator, orbital ordering, quantum spin-liquid, metal-insulator transition, multiferroicity, etc. [1–8] Because of the ground state instability due to such competing effects, a small external perturbation, e.g., doping, pressure, magnetic field, can readily modify the electronic and magnetic correlations and thus the ground state properties of systems even in the same family. [3,9–11] For example, among the ARuO$_3$ (A= Ca, Sr, Ba) Ruthenate family, SrRuO$_3$ with orthorhombic perovskite structure is an itinerant ferromagnet below 165 K, [12] whereas the iso-structural compound CaRuO$_3$ is a paramagnetic non-Fermi-liquid metal. [13] On the other hand, the iso-chemical compound BaRuO$_3$, where the Ba-cation has larger ionic radius compared to Sr/Ca, crystallizes in hexagonal or rhombohedral perovskite structure (four(4H)-, six(6H), or nine(9R)-layered structures) that depends on the synthesis condition, and thus exhibits various physical properties. [14]

In particular, 9R-BaRuO$_3$ (space group R-3m), which consists of corner-sharing Ru$_3$O$_{12}$-trimers, shows a metal-insulator-type (semiconductor-insulator) transition around 110 K which is accompanied by a structural change, a feature distinct from hexagonal BaRuO$_3$. [15] No long-range magnetic ordering is observed for all BaRuO$_3$ compounds (Ru$^{+4}$, S = 1 in low-spin state). [15] In contrast, Ba$_4$Ru$_3$O$_{10}$, which crystalizes in orthorhombic structure (space group *Cmca*) and consists of chains of $Ru_3O_{10}$-trimers running along *c*-axis in a zig-zag manner [16–19], exhibits a metal-insulator-type (semiconductor-insulator) transition around 105 K that is accompanied by antiferromagnetic (AFM) ordering but no structural transition. In addition, $Ba_4LnRu_3O_{12}$ (Ln= La, Rare-earth), which has an average valence state of +4.33/Ru (when Ln=$Ln^{+3}$), is iso-structural to 9R-$BaRuO_3$ (Space group R-3m) and consists of similar $Ru_3O_{12}$-trimers, though the trimers connected through $LnO_6$-octahedra. Nevertheless, different from 9R-$BaRuO_3$, $Ba_4LnRu_3O_{12}$ does not exhibit any structural or metal-insulator-type transition and undergoes long-range magnetic ordering at low temperature. The long-range-ordering (LRO) in semiconducting $Ba_4LnRu_3O_{12}$ with Ln being magnetic rare-earth atom is understood to be triggered by the rare-earth magnetic ordering. Intriguingly, $Ba_4LaRu_3O_{12}$, where $Ru_3O_{12}$-trimers are connected via non-magnetic La-atom, also exhibits long-range magnetic ordering at ~ 6 K, which is in sharp contrast to 9R-$BaRuO_3$. Similarly, iso-structural compound $Ba_4NbRu_3O_{12}$ with a different valence state of Ru (+3.67/Ru for $Ru_3O_{12}$-trimers) behaves as a Mott-insulator and magnetically orders around 4 K with a strong geometrical frustration. Therefore, both valence state and local crystal environment of Ru-ion play an important role on the physical properties of Ruthenates even in the similar family. The small structural change/distortion can significantly modify magnetic correlation and spin-lattice coupling, which leads to vastly different ground states and intriguing physical properties.

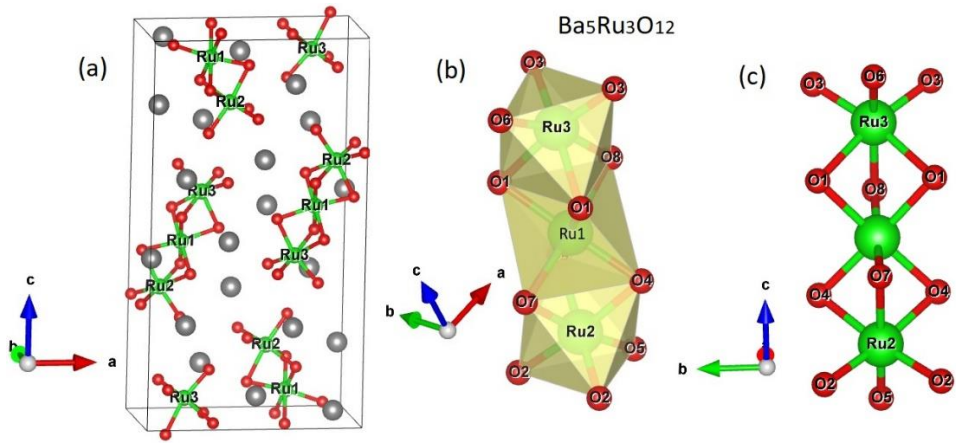

**Fig.1.** Crystal structure of $Ba_5Ru_3O_{12}$. The Ru-trimers are shown in (**b-c**).

The title compound, $Ba_5Ru_3O_{12}$ (an average valence state of +4.67/Ru), which has nearly similar crystal structure to the aforementioned 9R-$BaRuO_3$, $Ba_4Ru_3O_{10}$ or $Ba_4Ln(Nb)Ru_3O_{12}$, crystalizes in *Pnma* space group and consists of $Ru_3O_{12}$-trimers (Fig. 1). Each trimer is composed of face-sharing $RuO_6$-octahedra, similar to other Ruthenates mentioned above. Nevertheless, in contrast to other Barium Ruthenates, the $Ru_3O_{12}$-trimers of $Ba_5Ru_3O_{12}$ are not connected to each-other [16]. Interestingly, despite the isolated trimer structure, a previous report of magnetic susceptibility suggests that $Ba_5Ru_3O_{12}$ exhibits an AFM ordering below 60 K. [16] Despite that there are some studies of trimer Ruthenates where trimers are connected, thus far only a brief report exists on this isolated trimer system. [16] Therefore, it is highly desirable to investigate $Ba_5Ru_3O_{12}$ in detail to study the nature of its magnetic ground state and to better understand the electronic and magnetic correlation in this new trimer system.

In this paper we have reported comprehensive magnetic, transport, heat-capacity and neutron diffraction measurements on $Ba_5Ru_3O_{12}$. We have revealed a complex antiferromagnetic spin structure below $T_N \sim 60$ K which is presumably ascribed to combining effects of competing magnetic exchange interactions between Ru ions on different crystallographic sites and different orbital hybridization on different Ru sites. Neither structural phase transition nor a change in electronic properties is observed to accompany with the onset of magnetic ordering. A brief comparison to the magnetic structure of $Ba_4Ru_3O_{10}$ is also presented.

## II. Experimental Details

The polycrystalline $Ba_5Ru_3O_{12}$ sample was synthesized using solid state chemistry method by mixing high quality (>99.9%) chemical of $BaCO_3$ and $RuO_2$, as described in earlier report. [16] The stoichiometric mixture of raw materials was pressed into pellets and sintered in air at 600 ˚C for 24 h and then taken out to regrind. After repeating three cycles of this process, the powder was pressed into pellets and then sintered at 1200 ˚C for 24 h. Magnetic susceptibility measurements as a function of temperature and magnetic field were performed using a commercial SQUID-VSM magnetometer. The resistivity and heat capacity measurements were conducted using Physical Properties Measurements System (PPMS). Neutron powder diffraction measurements were carried out using a high-resolution time-of-flight neutron powder diffractometer (POWGEN) with a bandwidth with central wavelengths of 2.665 Å in Oak Ridge National Laboratory. An POWGEN automatic changer (PAC) was used to cover the temperature region of 10-300 K. The magnetic structure was resolved using Fullprof package and SARAh program. [20,21]

## III. Results

### A. Magnetic properties

The *DC* magnetic susceptibility ($\chi = M/H$) as a function of temperature in the presence of 1 kOe magnetic field is shown in Fig.2(a). The paramagnetic Curie-Weiss (C-W) behavior deviates below 120 K, implying the development of short-range correlation. With further lowering the temperature $\chi$ starts to decrease sharply below $\sim 60$ K ($T_N$), manifesting a well-defined long-range AFM ordering. The C-W fit ($\chi = \chi_0 + c/(T-\Theta_{CW})$) between 200-350 K yields Curie temperature ($\Theta_{CW}$) of -118 K and an effective magnetic moment ($\mu_{eff}$) of 4.42 $\mu_B$ per formula unit, with negligible $\chi_0$ = -0.00013 emu/mol. The negative value of $\Theta_{CW}$ indicates dominant AFM interactions in this system. The $\Theta_{CW}$ is higher than $T_N$, suggesting magnetic frustration with a frustration parameter ($|\Theta_{CW} / T_N|$) $\sim 2$. The isothermal magnetization ($M(H)$) below $T_N$ (as shown in Fig.2(b) for $T = 3$ and 40 K) exhibit a linear behavior as a function of magnetic field, supporting the AFM nature of this compound.

### B. Heat-capacity and Resistivity

The temperature dependent heat-capacity (*C*) measured at $H = 0$ and 50 kOe are plotted in Fig.2(c). The peak below $T_N$ confirms the magnetic ordering of this compound. We do not observe any appreciable change at $H = 50$ kOe, suggesting that the associated Zeeman energy is much smaller than the dominant magnetic exchange interaction. It is likely that there is already entropy loss at higher temperature as a result of short-range magnetic correlation from trimer. Thus, the change in entropy loss is low around the magnetic ordering temperature, which yields a small peak in the heat capacity. We have also measured *DC* resistivity for the titled compound. The resistivity increases exponentially with lowering the temperature down to 10 K (see inset of Fig. 2(a)), which indicates insulating behavior (with an activation energy of 0.05 eV) of this compound. An insulating behavior is also reported in $Ba_4Ln(Nb)Ru_3O_{12}$. [4] No metal-insulator transition is observed, unlike $BaRuO_3$ or $Ba_4Ru_3O_{10}$. [15,18]

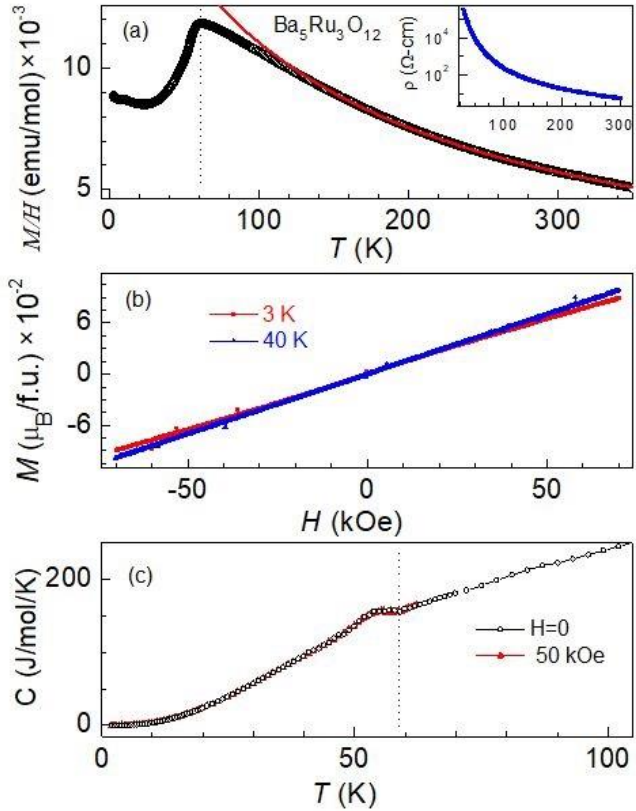

**Fig.2. (a)** The temperature depended *DC* magnetic susceptibility measured with 1 kOe magnetic field. The red curve is the Curie-Weiss fit in paramagnetic region from 200 to 350 K. The inset shows the resistivity as a function of temperature measured at zero magnetic field. **(b)** Isothermal magnetization as a function of magnetic field at $T = 3$ and 40 K. **(c)** Heat Capacity as a function of temperature measured at $H = 0$ and 5 T.

### C. Neutron Powder Diffraction

We have performed neutron power diffraction measurements to resolve the magnetic structure of $Ba_5Ru_3O_{12}$. The neutron diffraction profile measured at 100 K (above $T_N$) is well fitted with the reported space group *Pnma*, as shown in Fig.3(a), which affirms single-phase of the material. For $Ba_5Ru_3O_{12}$ there are three distinct inequivalent Ru-sites in $Ru_3O_{12}$-trimer (i.e., Ru1, Ru2, Ru3 (Fig.1)) but with same Wyckoff position 4c (x, 1/4, z). This is distinct from other trimer systems discussed above which have only two inequivalent Ru-sites with different Wyckoff positions. The atomic position of Ru1 (the middle Ru atom of the trimer), Ru2 and Ru3 in $Ba_5Ru_3O_{12}$ are (0.7859, 0.25, 0.5575), (0.8806, 0.25, 0.6735), (0.6935, 0.25, 0.4309), respectively (Fig.1). It is worth noting that the bond-length between Ru1 and Ru2 is ~2.51 Å, whereas it is ~2.69

Å between Ru1 and Ru3. $RuO_6$ octahedra are slightly distorted and the distortion is different for different Ru-sites. The bond angles of O-Ru-O are tabulated in Table-I.

**Table-I: The O-Ru-O bond of $RuO_6$ octahedron angle for different Ru and Oxygen sites for $Ba_5Ru_3O_{12}$.**

| < O-Ru1-O | | < O-Ru2-O | | < O-Ru3-O | |
|---|---|---|---|---|---|
| < O8-Ru1-O1 | $82.9^0$ | < O7-Ru2-O4 | $84.0^0$ | < O8-Ru3-O1 | $79.4^0$ |
| < O1-Ru1-O4 | $93.7^0$ | < O4-Ru2-O2 | $92.8^0$ | < O1-Ru3-O3 | $98.4^0$ |
| < O4-Ru1-O7 | $88.7^0$ | < O2-Ru2-O5 | $94.5^0$ | < O3-Ru3-O6 | $95.8^0$ |
| < O7-Ru1-O8 | $173.4^0$ | < O5-Ru2-O7 | $172.7^0$ | < O6-Ru3-O8 | $164.9^0$ |
| < O1-Ru1-O4 | $178.3^0$ | < O2-Ru2-O4 | $173.8^0$ | < O1-Ru3-O3 | $171.1^0$ |

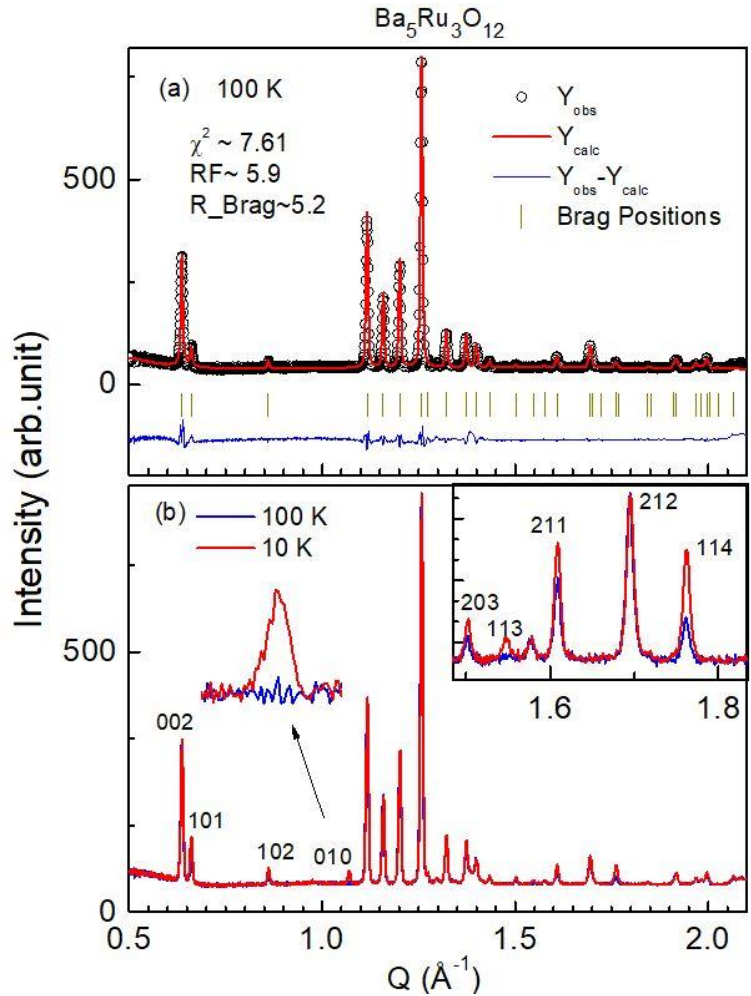

**Fig.3. (a)** Rietveld refinement to the neutron powder diffraction pattern measured at $T = 100$ K. The open black circle represents the experimental data, while the red solid line shows the Rietveld fitting. The vertical bars display the Bragg peak positions of crystal structure. The continuous blue line at the bottom shows the difference between the experimental and calculated intensity. **(b)** Comparison of neutron powder diffraction pattern collected at $T = 10$ and 100 K. The inset shows an expanded view at higher Q. The (H K L) values are indexed for most of reflections.

The neutron diffraction pattern as a function of the momentum transfer (Q) measured at $T = 10$ and 100 K is plotted in Fig. 3(b), which shows the change of Bragg peak intensity below and above $T_N$. At $T = 10$ K, besides the enhanced intensity of some of Bragg reflections compared to the 100 K data, an additional Bragg peak is observed at Q ~ 1.07 Å$^{-1}$ that corresponds to (0 1 0). Furthermore, no structural change is observed at 10 K compared to that of 100K, which excludes any structural phase transition accompanying with the magnetic transition.

**Table-II:** The irreducible representations and basis vectors for each Ru-site (4c) for space group *Pnma* and the propagation vector $\mathbf{k} = (0\ 0\ 0)$.

| I.R. | B.V. | x, y, z | | | x+1/2, -y+1/2, -z+1/2 | | | -x, y+1/2, -z | | | -x+1/2, -y, z+1/2 | | |
|---|---|---|---|---|---|---|---|---|---|---|---|---|---|
| | | $m_a$ | $m_b$ | $m_c$ | $m_a$ | $m_b$ | $m_c$ | $m_a$ | $m_b$ | $m_c$ | $m_a$ | $m_b$ | $m_c$ |
| $\Gamma_1$ | $\Psi_1$ | 0 | 2 | 0 | 0 | -2 | 0 | 0 | 2 | 0 | 0 | -2 | 0 |
| $\Gamma_2$ | $\Psi_2$ | 2 | 0 | 0 | 2 | 0 | 0 | -2 | 0 | 0 | -2 | 0 | 0 |
| | $\Psi_3$ | 0 | 0 | 2 | 0 | 0 | -2 | 0 | 0 | -2 | 0 | 0 | 2 |
| $\Gamma_3$ | $\Psi_4$ | 2 | 0 | 0 | 2 | 0 | 0 | 2 | 0 | 0 | 2 | 0 | 0 |
| | $\Psi_5$ | 0 | 0 | 2 | 0 | 0 | -2 | 0 | 0 | 2 | 0 | 0 | -2 |
| $\Gamma_4$ | $\Psi_6$ | 0 | 2 | 0 | 0 | -2 | 0 | 0 | -2 | 0 | 0 | 2 | 0 |
| $\Gamma_5$ | $\Psi_7$ | 0 | 2 | 0 | 0 | 2 | 0 | 0 | 2 | 0 | 0 | 2 | 0 |
| $\Gamma_6$ | $\Psi_8$ | 2 | 0 | 0 | -2 | 0 | 0 | -2 | 0 | 0 | 2 | 0 | 0 |
| | $\Psi_9$ | 0 | 0 | 2 | 0 | 0 | 2 | 0 | 0 | -2 | 0 | 0 | -2 |
| $\Gamma_7$ | $\Psi_{10}$ | 2 | 0 | 0 | -2 | 0 | 0 | 2 | 0 | 0 | -2 | 0 | 0 |
| | $\Psi_{11}$ | 0 | 0 | 2 | 0 | 0 | 2 | 0 | 0 | 2 | 0 | 0 | 2 |
| $\Gamma_8$ | $\Psi_{12}$ | 0 | 2 | 0 | 0 | 2 | 0 | 0 | -2 | 0 | 0 | -2 | 0 |

The propagation vector is found to be $\mathbf{k} = (0\ 0\ 0)$. The irreducible representations (I.R.) and basis vectors (B.V.) of the Ru1 spins for *Pnma* space group associated with the propagation vector, obtained from SARAh program, are shown in Table II. Because of the same Wyckoff site of three Ru-atoms, obviously, Ru2 and Ru3-atoms have exactly the same I.R. and B.V as Ru1-atom. There are eight irreducible representations for each Ru-atom, as represented by $\Gamma_{mag}$ (Ru)= $1\Gamma_1^1 + 2\Gamma_2^1 + 2\Gamma_3^1 + 1\Gamma_4^1 + 1\Gamma_5^1 + 2\Gamma_6^1 + 2\Gamma_7^1 + 1\Gamma_8^1$. The magnetic moments along crystallographic *a*, *b* and *c*-axis are described by $\Psi_2(\Gamma_2)$, $\Psi_4(\Gamma_3)$, $\Psi_8(\Gamma_6)$, $\Psi_{10}(\Gamma_7)$ B.V., $\Psi_1(\Gamma_1)$, $\Psi_6(\Gamma_4)$, $\Psi_7(\Gamma_5)$, $\Psi_{12}(\Gamma_8)$

B.V., and $\Psi_3(\Gamma_2)$, $\Psi_5(\Gamma_3)$, $\Psi_9(\Gamma_6)$, $\Psi_{11}(\Gamma_7)$ B.V., respectively. Since neutrons couple to the magnetic moment component perpendicular to **Q**, the absence of noticeable difference in the (0 0 2) Bragg peak intensity measured at 10 K and 100 K, suggests that the magnetic moment is oriented along *c*, or at least the *c*-component of magnetic moment ($M_c$) is dominant. First, we consider the basis functions which only have $M_c$ component, that is, $\Psi_3(\Gamma_2)$, $\Psi_5(\Gamma_3)$, $\Psi_9(\Gamma_6)$, and $\Psi_{11}(\Gamma_7)$. Among these basis functions, $\Psi_9(\Gamma_6)$ yields best profile matching in Rietveld refinement, as shown in Fig.4(a).

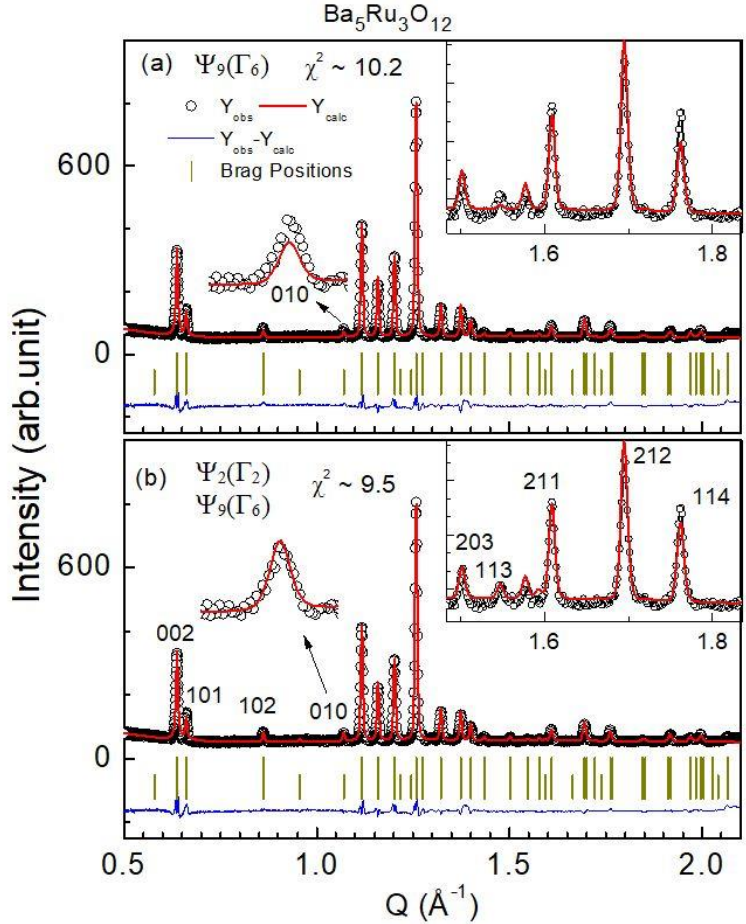

**Fig.4.** Rietveld refinement to the neutron powder diffraction pattern measured at *T* = 10 K modeled with **(a)** $\Psi_9(\Gamma_6)$ only and **(b)** a combination of $\Psi_2(\Gamma_2)$ and $\Psi_9(\Gamma_6)$. The open black circles represent the experimental data, while the red solid curve shows the Rietveld fitting. The vertical bars display the Bragg peak positions of crystal structure, the next lower vertical lines represent magnetic Bragg peaks associated with the propagation vector ***k*** = (0 0 0). The continuous blue line at the bottom of the figure shows the difference between the experimental and calculated intensity. The insets show an expanded view of some Bragg reflections.

The obtained magnetic structure is illustrated in Fig.5(a) where magnetic moments are collinearly aligned along the *c*-axis. Ru1 and Ru2 spins are parallel aligned, while Ru1 and Ru3 spins are antiparallel aligned. This is reasonable considering FM direct exchange interaction between Ru1 and Ru2 due to metallic bond length (less than Ru-Ru bond ~2.65 Å in Ru-metal) and the antiferromagnetic super-exchange interaction of Ru1-O-Ru3 bond. The FM direct exchange interaction between Ru1 and Ru2 is expected to compete with the Ru1-O-Ru2 AFM super-exchange interaction, giving rise to magnetic frustration. We notice that although the refinement using $\Psi_9(\Gamma_6)$ basis function nearly captures the extra magnetic Bragg peak (0 1 0) and the enhanced Bragg peak intensity of most of nuclear reflections, there is a mismatch in experimentally and theoretically obtained intensity of the (1 1 4) reflection (see Inset of Fig 4(a)). In addition, the negligible change in the intensity of (2 0 0) Bragg peak between 10 K and 100 K suggests the possibility of *a*-component of magnetic moment ($M_a$). Thus, we have tried a combination of basis functions which can give magnetic moment in *ac*-plane. We find that the low temperature neutron scattering data is best fitted using a combination of $\Psi_2(\Gamma_2)$ and $\Psi_9(\Gamma_6)$, as shown in Fig.4(b). For instance, as shown in the inset of Fig. 4(b), both (0 1 0) and (1 1 4) Bragg peaks are much better fitted using this model compared to those using the other earlier one (see the inset of Fig 4(a)). The obtained magnetic structure based on this refinement is depicted in Fig.5(b). The magnetic moments are non-collinear and confined in the *ac*-plane. The moment of Ru1 (middle of trimer) are oriented along c-axis, and the moments of Ru2 and Ru3 are oriented in ac-plane with a canting angle of 74.4° and 62° relative to the *c*-axis. The moment size of Ru1, Ru2, Ru3 is ~1.52, 1.36, and 0.91 $\mu_B$, respectively. This yields a total moment of 3.79 $\mu_B$ per $Ru_3O_{12}$-trimer.

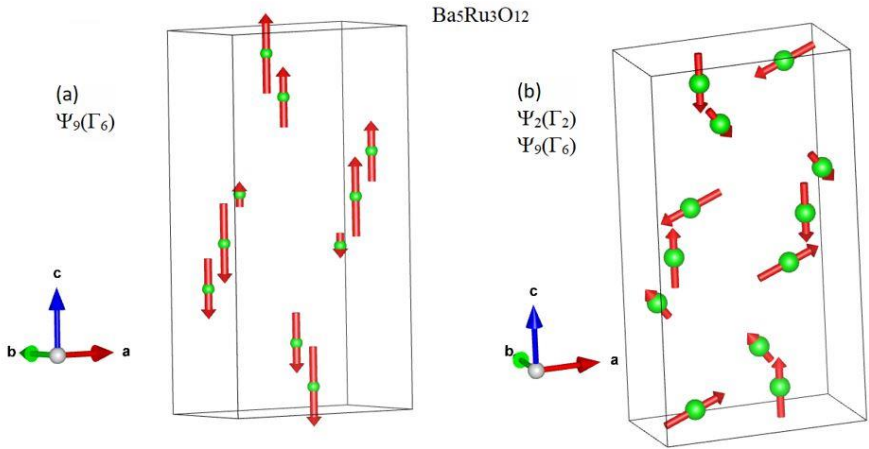

**Fig.5.** Magnetic structure of $Ba_5Ru_3O_{12}$ modeled with **(a)** $\Psi_9(\Gamma_6)$ only and **(b)** a combination of $\Psi_2(\Gamma_2)$ and $\Psi_9(\Gamma_6)$.

Such a canted magnetic structure possibly arises from different competing nearest neighbor and next-nearest neighbor interactions between different inequivalent Ru-sites. Based on Goodenough-Anderson-Kanamori rules, the nearest-neighbor exchange interactions ($J_{nn}$) includes i) Ru1-Ru2 ferromagnetic direct exchange interaction, ii) Ru1-O-Ru2 AFM super-exchange interaction, and iii) Ru1-O-Ru3 AFM super-exchange interaction. However, the next-nearest-neighbor super-super-exchange interactions ($J_{nnn}$) between Ru1 and Ru3 in the trimer may not be negligible. The competition among $J_{nn}$ and $J_{nnn}$, where $J_{nnn} < J_{nn}$, introduces exchange frustration and therefore may stabilize the system with a canted spin structure. The neighboring trimers are antiferromagnetically coupled via the super-super-exchange interaction (i.e. Ru1-O-O-Ru1), which yields 3D long-range AFM ordering.

A temperature- and *d*-dependent neutron scattering intensity map is shown in Fig.6. One can see that the (0 1 0) reflection emerges and the intensity of (1 0 2) becomes enhanced below 60 K, which further confirms the magnetic ordering.

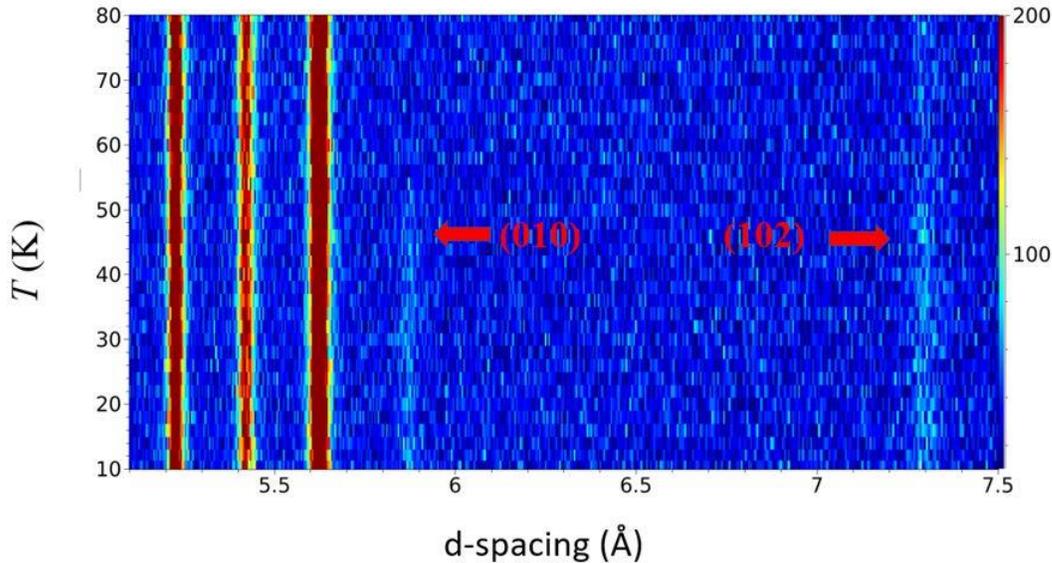

**Fig.6.** Temperature- and *d*-dependent neutron scattering intensity map. The red arrows point to the onset of magnetic transition.

We have also performed the neutron scattering measurements on another trimer Ruthenate $Ba_4Ru_3O_{10}$ for comparison, which is documented in the Supplemental Material. [22] The good fitting of Rietveld refinement at 140 K, ( Fig.S2a in the Supplemental Material [22] ), confirms the desired structure with *Cmca* space group as reported earlier. [18] The Rietveld refinement at 10 K is depicted in Fig.S2b in the Supplemental Material [22]. A preliminary neutron diffraction study by Klein et al. [18] documented an enhancement on (002)-peak below magnetic ordering and proposed a magnetic structure. In addition to (002) Bragg peak, we have observed an enhancement of the neutron diffraction intensity of some other additional Bragg peaks compared to the data taken at 140 K. Some of those magnetic Bragg peaks are depicted in Insets of Fig. S2a in the Supplemental Material [22]. All these peaks are well-modeled with propagation vector *k* = (0 0 0) (see Fig. S2b and Insets in the Supplemental Material [22]). The magnetic structure obtained from the refinement confirms the prediction of earlier report by Klein et al. [18] There is no magnetic moment on Ru1, thereby the trimer essentially behaves like a dimer, which is distinct from $Ba_5Ru_3O_{12}$. The moment on Ru2 is 1.05 $\mu_B$ and points to the *b*-direction. The spins on Ru2 are antiferromagnetically coupled within a trimer. The spins in two adjacent trimers are ferromagnetically coupled along the *a*-direction and antiferromagnetically coupled along the *c*-direction (See Fig. S3 in Supplemental Material [22] ).

## IV. Discussion and Conclusion

As discussed previously, the obtained magnetic moment of $Ba_5Ru_3O_{12}$ based on Rietveld refinement to the neutron diffraction data is about 3.79 $\mu_B$ per $Ru_3O_{12}$-trimer. For the $RuO_6$ octahedron, the $d$-orbital splits into lower energy $t_{2g}$ and higher energy $e_g$ orbitals due to crystal field effect in octahedral symmetry. Thus, considering discrete Ru-atom, naively one would expect that $Ru^{+4}$ ($d^4$) has four electrons in $t_{2g}$-orbital which would yield $S = 1$ effective quantum number and that $Ru^{+5}$ ($d^3$) yields $S = 3/2$ effective quantum number [See Fig.7(a)]. This would give a total moment of 8 $\mu_B$ (considering Lande-$g$ factor of 2) for three Ru-atom, i.e. per $Ru_3O_{12}$-trimer, which is much higher than the experimentally obtained moment of $Ru_3O_{12}$-trimer. Therefore, other mechanisms need to be considered to account for the reduced magnetic moment in this system.

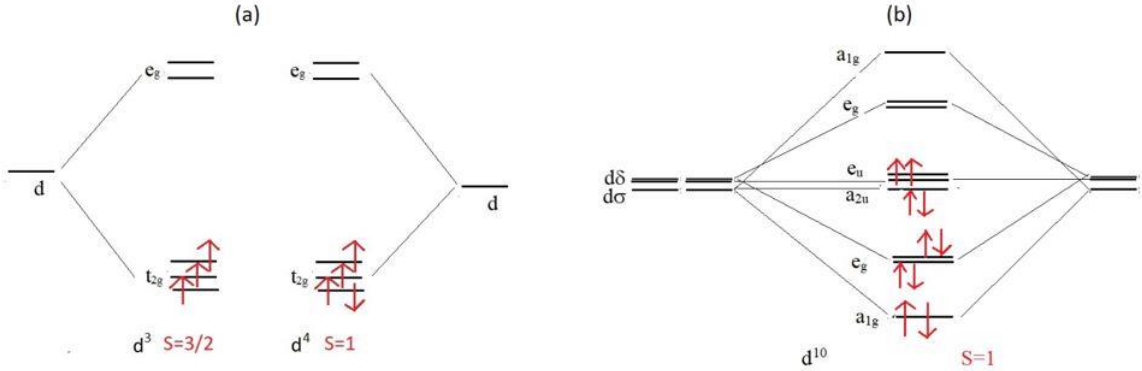

**Fig.7.** (a) The electronic configuration of $t_{2g}$-$e_g$-model for $d^3$ and $d^5$, and (b) electronic configuration of Ru-trimer in $D_{3d}$ symmetry due to metal-metal orbital hybridizations.

First, we consider molecular orbital of $Ru_3O_{12}$ trimer with $D_{3d}$ symmetry due to Ru-Ru metallic bonding, as proposed for some of other trimer systems. [4] The 9R-$BaRuO_3$ compound does not exhibit any magnetic ordering. The Ru-Ru distance is ~2.53 Å which is shorter than Ru-metal, yielding a hybridization between $d$-orbitals. [23] Because of this strong metal-metal bonding of Ru atoms within a trimer, the whole trimer may behave as a single molecular-orbital-like state instead of three discrete Ru-orbitals, if the kinetic energy gain due to metallicity is larger than coulomb interaction (Hund's coupling). The total valence electron count of $Ru_3O_{12}$ trimer is 12 ($Ru^{+4}$ → $d^4$). Considering the metal-metal bonding, the electronic configuration of a $Ru_3O_{12}$-trimer with $D_{3d}$ symmetry [24] is $(a_{1g})^2(e_g)^4(a_{2u})^2(e_u)^4$, which yields $S = 0$, and thus the compound behaves as a non-magnet. [18] Similarly, metal-metal bonding with single molecular-like-orbital state has been predicted in another trimer compound $Ba_4NbRu_3O_{12}$, which consist of 13-electron in effective $d$-orbital of $Ru_3O_{12}$ trimer and thus yields an effective $S = 1/2$ in the trimer. [4] However, the same picture cannot be applied to $Ba_4Ru_3O_{10}$ ($Ru^{+4}$ → $d^4$) which exhibits magnetic ordering at 105 K with non-zero spin-moment at Ru2-site of the trimer. And for $Ba_5Ru_3O_{12}$, it contains 10 valence electrons in total for each $Ru_3O_{12}$ trimer. As a result, considering $D_{3d}$ symmetry of $Ru_3O_{12}$-trimer, [24] that is, $(a_{1g})^2(e_g)^4(a_{2u})^2(e_u)^2$, one would anticipate $S = 1$ in its ground state [Fig.7(b)]. The magnetic moment obtained based on this model is smaller than the experimental value of this compound.

Second, we consider the effects of spin-orbit coupling. It is known that, in octahedral symmetry, the spin-orbit coupling ($\lambda$) may split the Ru orbital triplet $^3T_{1g}$ into three sublevels with energies E = $-2\lambda$ ($J = 0$), E = $-\lambda$ ($J = 1$), and E = $\lambda$ ($J = 2$). In this case, $Ru^{4+}$ cations adopt $J = 0$ in its ground state and therefore the system would not order magnetically. [18] However, if the $RuO_6$ octahedra in trimers are not symmetrically connected in all direction, it may lift the degeneracy of the $t_{2g}$ orbitals and thus leads to a ground state with non-zero magnetic moment. If we consider three discrete Ru-orbital instead of a single trimer-orbital picture, this model can account for the difference in the magnetic ground states of $BaRuO_3$ and $Ba_4Ru_3O_{10}$. For $BaRuO_3$, the distortion of $RuO_6$ octahedron is small with the bond angle for O-Ru-O ~ $180^0$, thus, the nonmagnetic $J = 0$ state can naturally apply to $BaRuO_3$. In contrast, in $Ba_4Ru_3O_{10}$ the $RuO_6$ octahedron of Ru2 atom is slightly distorted with ~$171^0$ bond angle for the four in-plane O-Ru2-O bonds and ~$180^0$ for the rest two O-Ru2-O bonds with apical oxygen atoms. It is hypothetically argued [18] that the perturbation of octahedral symmetry of Ru2 is large enough to lift the degeneracy of $t_{2g}$-orbital but not for the Ru1. This results in non-zero moment of the ground state of two Ru-atoms at edge of the trimer (Ru2) but zero moment of the center Ru atom of the trimer (Ru1). For $Ba_5Ru_3O_{12}$, there are three distinct Ru-atoms, namely, Ru1, Ru2 and Ru3 and the average valence state is +4.67 (+5 for two Ru and +4 for one Ru). The Ru1-O distances are 1.98 - 2.0 Å, whereas, the Ru2-O and Ru3-O distances are 1.88 - 2.07 Å and 1.86 - 2.1 Å. The Ru-O-Ru angle for different Ru-atoms are presented in Table-I. Considering no spin-orbit coupling for $Ru^{+5}$ (zero orbital degrees of freedom) and finite spin-orbit coupling for $Ru^{+4}$, the total moment of trimer (S=3/2 for two Ru and J=0 for one Ru) will be 6 $\mu_B$, which is higher than the experimentally obtained value. Also, for $Ba_5Ru_3O_{12}$, as tabulated in in Table-I, there is large deviation of the O-Ru-O bond angles from $180^0$, as observed in Ru2-atom for $Ba_4Ru_3O_{10}$, which also indicates that probably, this model (spin-orbit coupling) alone is not valid for our title compound.

Thus, the above two mechanisms that have been proposed for other trimers systems could not be applied individually to account for the experimentally obtained effective moment in $Ba_5Ru_3O_{12}$. It is likely that $J_{nn}$ and $J_{nnn}$ are comparable to each-other and thus the competition between nearest and next-nearest exchange interaction plays a crucial role for geometrical frustration, unlike $BaRuO_3$ and $Ba_4Ru_3O_{10}$-trimer systems. On the other hand, the different moment size of Ru on different crystallographic site indicates that orbital hybridization or/and spin-orbit coupling could be different for different Ru atoms. Our results suggest that the magnetism in this compound is quite complex, on which the combined effects of different factors arising from geometrical frustration, spin-orbit coupling, and orbital hybridization determine the magnetic ground state. Further different spectroscopic and theoretical investigations are warranted to shed more light on this fascinating aspect.

In summary, our detailed investigation on $Ba_5Ru_3O_{12}$ trimer system reveals a long-range antiferromagnetic ordering below 60 K. A complex magnetism with canted spin-structure is observed, distinct from other trimer systems such as $BaRuO_3$, $Ba_4Ru_3O_{10}$, or, $Ba_4LnRu_3O_{12}$. Such a complex behavior arises as a result of a combination of different level of hybridization (localization) on different Ru sites and strong spin-frustration in this trimer system. The system exhibits insulating behavior throughout the temperature range measured. No metal-insulator-like transition is observed, unlike $BaRuO_3$ and $Ba_4Ru_3O_{10}$. This study demonstrates that the valence state and hybridization of Ru-atom, together with the $RuO_6$ octahedral distortion, play an important role on electronic and magnetic correlations in Ruthenates.

# Supplemental Material

## Neutron Powder Diffraction of $Ba_4Ru_3O_{10}$

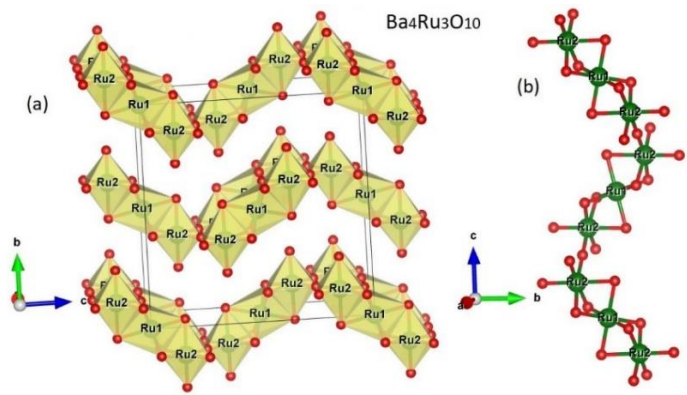

**Fig. S1.** Crystal structure of $Ba_4Ru_3O_{10}$.

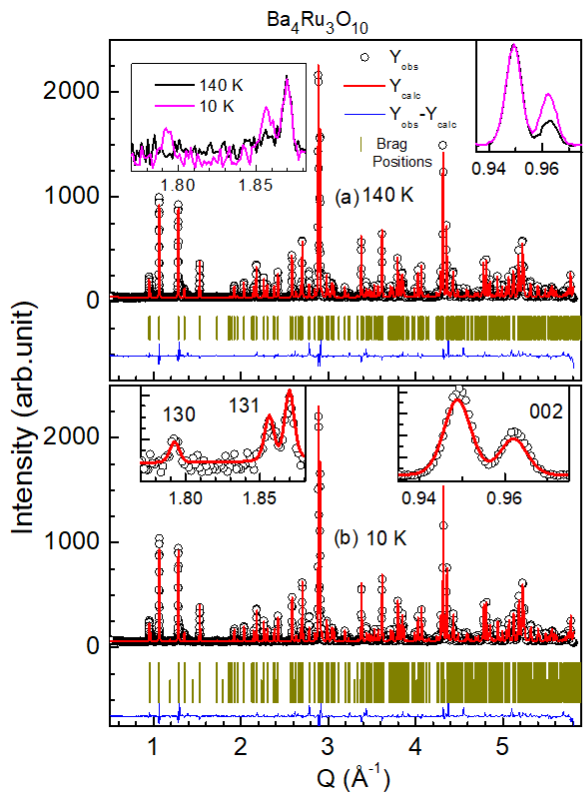

**Fig. S2.** Rietveld fitting of *t.o.f* powder neutron diffraction pattern collected at (a)140 K and (b) 10 K for the compound $Ba_4Ru_3O_{10}$. The open black circle represents the experimental data, while the red solid line shows the Rietveld fitting. The vertical bars display the Bragg peak positions of crystal structure of $Ba_5Ru_3O_{12}$, the next lower vertical lines represent magnetic Bragg peaks associated with $k = (0\ 0\ 0)$. The continuous blue line at the bottom of the figure shows the difference between the experimental and calculated intensity. Insets shows the magnified picture of some (H K L) peaks.

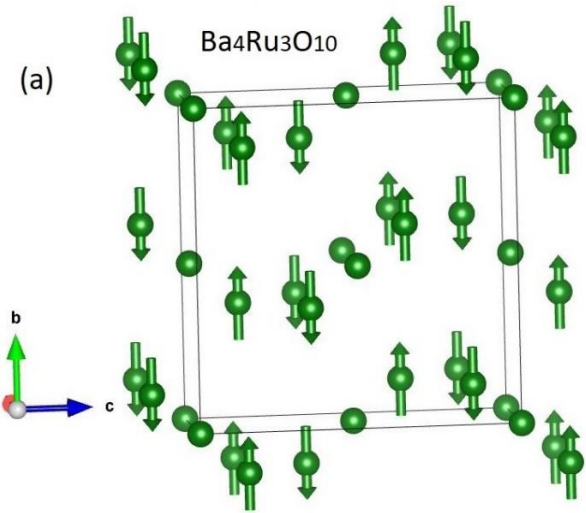

**Fig. S3.** Magnetic structure of $Ba_4Ru_3O_{10}$.

**Acknowledgement:**

Work at Michigan State University was supported by the U.S. Department of Energy, Office of Science, Office of Basic Energy Sciences, Materials Sciences and Engineering Division under Award No. DE-SC0019259. A portion of this research used resources at Spallation Neutron Source, a DOE Office of Science User Facility operated by the Oak Ridge National Laboratory.

**References:**

[1] Y. Maeno, H. Hashimoto, K. Yoshida, S. Nishizaki, T. Fujita, J. G. Bednorz, and F. Lichtenberg, *Superconductivity in a Layered Perovskite without Copper*, Nature **372**, 532 (1994).
[2] A. Banerjee, C. A. Bridges, J.-Q. Yan, A. A. Aczel, L. Li, M. B. Stone, G. E. Granroth, M. D. Lumsden, Y. Yiu, J. Knolle, S. Bhattacharjee, D. L. Kovrizhin, R. Moessner, D. A. Tennant, D. G. Mandrus, and S. E. Nagler, *Proximate Kitaev Quantum Spin Liquid Behaviour in a Honeycomb Magnet*, Nature Materials **15**, 733 (2016).
[3] J. Leshen, M. Kavai, I. Giannakis, Y. Kaneko, Y. Tokura, S. Mukherjee, W.-C. Lee, and P. Aynajian, *Emergent Charge Order near the Doping-Induced Mott-Insulating Quantum Phase Transition in Sr 3 Ru 2 O 7*, Communications Physics **2**, 1 (2019).
[4] L. T. Nguyen, T. Halloran, W. Xie, T. Kong, C. L. Broholm, and R. J. Cava, *Geometrically Frustrated Trimer-Based Mott Insulator*, Phys. Rev. Materials **2**, 054414 (2018).
[5] R. J. Cava, *Schizophrenic Electrons in Ruthenium-Based Oxides*, Dalton Trans. 2979 (2004).
[6] P. Khalifah, R. Osborn, Q. Huang, H. W. Zandbergen, R. Jin, Y. Liu, D. Mandrus, and R. J. Cava, *Orbital Ordering Transition in La4Ru2O10*, Science **297**, 2237 (2002).
[7] T. Basu, V. Caignaert, S. Ghara, X. Ke, A. Pautrat, S. Krohns, A. Loidl, and B. Raveau, *Enhancement of Magnetodielectric Coupling in 6H-Perovskites Ba3RRu2O9 for Heavier Rare-Earth Cations (R = Ho, Tb)*, Phys. Rev. Mater. **3**, 114401 (2019).

[8] T. Basu, V. Caignaert, F. Damay, T. W. Heitmann, B. Raveau, and X. Ke, *Cooperative $\mathrm{Ru}(4d)\ensuremath{-}\mathrm{Ho}(4f)$ Magnetic Ordering and Phase Coexistence in the $6H$ Perovskite Multiferroic ${\mathrm{Ba}}_{3}\mathrm{Ho}{\mathrm{Ru}}_{2}{\mathrm{O}}_{9}$*, Phys. Rev. B **102**, 020409 (2020).

[9] M. Braden, O. Friedt, Y. Sidis, P. Bourges, M. Minakata, and Y. Maeno, *Incommensurate Magnetic Ordering in ${\mathrm{Sr}}_{2}{\mathrm{Ru}}_{1\ensuremath{-}\mathit{x}}{\mathrm{Ti}}_{\mathit{x}}{O}_{4}$*, Phys. Rev. Lett. **88**, 197002 (2002).

[10] K. Yoshida, F. Nakamura, T. Goko, T. Fujita, Y. Maeno, Y. Mori, and S. NishiZaki, *Electronic Crossover in the Highly Anisotropic Normal State of ${\mathrm{Sr}}_{2}{\mathrm{RuO}}_{4}$ from Pressure Effects on Electrical Resistivity*, Phys. Rev. B **58**, 15062 (1998).

[11] M. Zhu, J. Peng, W. Tian, T. Hong, Z. Q. Mao, and X. Ke, *Tuning the Competing Phases of Bilayer Ruthenate $\mathrm{C}{\mathrm{a}}_{3}\mathrm{R}{\mathrm{u}}_{2}{\mathrm{O}}_{7}$ via Dilute Mn Impurities and Magnetic Field*, Phys. Rev. B **95**, 144426 (2017).

[12] L. Klein, J. S. Dodge, C. H. Ahn, J. W. Reiner, L. Mieville, T. H. Geballe, M. R. Beasley, and A. Kapitulnik, *Transport and Magnetization in the Badly Metallic Itinerant Ferromagnet*, J. Phys.: Condens. Matter **8**, 10111 (1996).

[13] L. Klein, L. Antognazza, T. H. Geballe, M. R. Beasley, and A. Kapitulnik, *Possible Non-Fermi-Liquid Behavior of ${\mathrm{CaRuO}}_{3}$*, Phys. Rev. B **60**, 1448 (1999).

[14] J. G. Zhao, L. X. Yang, Y. Yu, F. Y. Li, R. C. Yu, Z. Fang, L. C. Chen, and C. Q. Jin, *Structural and Physical Properties of the 6H BaRuO3 Polymorph Synthesized under High Pressure*, Journal of Solid State Chemistry **180**, 2816 (2007).

[15] J. T. Rijssenbeek, R. Jin, Yu. Zadorozhny, Y. Liu, B. Batlogg, and R. J. Cava, *Electrical and Magnetic Properties of the Two Crystallographic Forms of BaRuO 3*, Phys. Rev. B **59**, 4561 (1999).

[16] C. Dussarrat, F. Grasset, R. Bontchev, and J. Darriet, *Crystal Structures and Magnetic Properties of Ba4Ru3O10 and Ba5Ru3O12*, Journal of Alloys and Compounds **233**, 15 (1996).

[17] T. Igarashi, Y. Nogami, Y. Klein, G. Rousse, R. Okazaki, H. Taniguchi, Y. Yasui, and I. Terasaki, *X-Ray Crystal Structure Analysis and Ru Valence of Ba4Ru3O10 Single Crystals*, J. Phys. Soc. Jpn. **82**, 104603 (2013).

[18] Y. Klein, G. Rousse, F. Damay, F. Porcher, G. André, and I. Terasaki, *Antiferromagnetic Order and Consequences on the Transport Properties of Ba${}_{4}$Ru${}_{3}${O}_{10}$*, Phys. Rev. B **84**, 054439 (2011).

[19] S. V. Streltsov and D. I. Khomskii, *Unconventional Magnetism as a Consequence of the Charge Disproportionation and the Molecular Orbital Formation in Ba${}_{4}$Ru${}_{3}$O${}_{10}$*, Phys. Rev. B **86**, 064429 (2012).

[20] J. Rodríguez-Carvajal, *Recent Advances in Magnetic Structure Determination by Neutron Powder Diffraction*, Physica B: Condensed Matter **192**, 55 (1993).

[21] A. S. Wills, *A New Protocol for the Determination of Magnetic Structures Using Simulated Annealing and Representational Analysis (SARAh)*, Physica B: Condensed Matter **276–278**, 680 (2000).

[22] *Supplemental Material at [URL Will Be Inserted by Publisher] for [Give Brief Description of Material]*, (n.d.).

[23] Y. A. Ying, Y. Liu, T. He, and R. J. Cava, *Magnetotransport Properties of BaRuO$_3$: Observation of Two Scattering Rates*, Phys. Rev. B **84**, 233104 (2011).

[24] B. E. Bursten, F. A. Cotton, and A. Fang, *Ground-State Electronic Structures and Other Electronic Properties of the Octahedral and Oligooctahedral Ruthenium Complexes Hexachlororuthenium(III), Nonachlorodiruthenium(III) and Dodecachlorotriruthenium(II, 2III)*, Inorg. Chem. **22**, 2127 (1983).